\documentclass[prb,twocolumn,superscriptaddress,showpacs,amsmath,amssymb]{revtex4}

\usepackage{graphicx}
\usepackage{latexsym}
\usepackage{amsmath}
\usepackage{amssymb}
\usepackage{amsfonts}
\usepackage{color}

\newcommand{\ds}{\displaystyle}

\begin{document}

\title{Little-Parks Oscillations in Hybrid Ferromagnet-Superconductor Systems}

\author{A.~V.~Samokhvalov}
\affiliation{Institute for Physics of Microstructures,
Russian Academy of Sciences, 603950, Nizhny Novgorod, Russia}
\author{A.~S.~Mel'nikov}
\affiliation{Institute for Physics of Microstructures,
Russian Academy of Sciences, 603950, Nizhny Novgorod, Russia}
\author{J-P.~Ader}
\affiliation{Institut Universitaire de France and Universit\'e Bordeaux I;
CNRS; CPMOH, F-33405 Talence, France}
\author{A.~I.~Buzdin}
\affiliation{Institut Universitaire de France and Universit\'e Bordeaux I;
CNRS; CPMOH, F-33405 Talence, France}

\date{\today}
\begin{abstract}
On the basis of of linearized Usadel equations we consider
superconductivity nucleation in multiply connected mesoscopic
superconductor/ferromagnet hybrids such as thin-walled superconducting
cylinders placed in electrical contact with a ferromagnetic metal.
We study the interplay between the oscillations of $T_c$ due to the
Little--Parks effect and the oscillations due to the exchange field.
We demonstrate that the exchange field provokes the switching
between the superconducting states with different vorticities
and this may result in the increase the critical temperature of the
superconducting transition in the magnetic field.
Moreover we analyse the influence of the S/F transparency on the
realisation of the states with higher vorticities.
\end{abstract}

\pacs{%
74.25.Dw, 
74.45.+c, 
74.78.Na} 

\maketitle

\section{Introduction}

Little-Parks effect,
\cite{Little-Parks-PR64}
i.e., the oscillations of the critical temperature $T_{c}$
of multiply-connected superconducting samples
in an applied magnetic field $H$,
is one of the striking phenomena demonstrating
coherent nature of the superconducting state.
Nevertheless, such oscillatory phase transition line $T_c(H)$
was shown to be inherent to simply connected
mesoscopic samples with the lateral size of the order
of several coherence lengths $\xi$
\cite{Buisson-PL90,Bruyndoncx-PRB99,Jadallah-PRL99}
and to hybrid ferromagnet (F)-superconductor (S) systems
with magnetic dots or domains
\cite{Lange-PRL03,Aladyshkin-JCM03,Aladyshkin-PRB03,Schildermans-PRB08},
that create a "magnetic template" for nucleation of the
superconducting order parameter.
These $T_c(H)$ oscillations reflect the switching
between vortex states characterized by different
winding numbers and are associated with the orbital effect
\cite{Ginzburg-JETP56}.

Another mechanism of switching between the superconducting
states with different vorticities in multiply connected
hybrid S/F structures such as thin-walled superconducting
shell placed in electrical contact with a ferromagnetic
cylinder was suggested recently in
\cite{Samokhvalov-PRB07}.
This mechanism is caused by the exchange interaction
and associated with the damped-oscillatory behavior
of the Cooper pair wave function in a ferromagnet
\cite{Buzdin-RMP05}.
It was shown that under certain conditions the exchange interaction
can stimulate the superconducting states with a nonzero
vorticity in the absence of external magnetic field.
The interplay between the exchange and orbital effects
may result in a subsequent switching between the
states with different vorticities, as the F core radius
increases. An obvious consequence of
these transitions between the states with different $L$
should be a nonmonotonic dependence of
the critical temperature $T_c$
on the F core radius and exchange field.
Similar oscillating behavior of $T_c$ on a ferromagnetic layer thickness
has been predicted for layered S/F structures \cite{Buzdin-JETPL90,Radovic-PRB91}.
Note, that the unusual ground states with spontaneously formed vortices in multiply-connected
S/F hybrids resemble the behavior of Josephson $\pi$-junction
\cite{Buzdin-JETPL82,Buzdin-JETPL91,Ryazanov-PRL01} with a step-like change
in the F layer thickness \cite{Bulaevsky-SSC78,Goldobin-PRL06,Frolov-PRB06}.

It is the purpose of this paper to study the influence
of an external magnetic field on the proximity induced
switching between the vortex states. We focus on the behavior of critical temperatures
for superconducting states with different vorticities
and, thus, we study the Little-Parks effect affected by the
exchange interaction.

The paper is organized as follows.
In Sec.~\ref{ModelSection} we briefly discuss the basic equations.
In Sec.~\ref{ResultsSection} we study the switching between
different vortex states for two model S/F systems placed in the external
magnetic field.
The first system consists of a thin-walled superconducting
cylindrical shell surrounding a cylinder of
a ferromagnetic metal.
As a second example we consider a cylindrical cavity in a bulk ferromagnet covered
by a thin layer of superconducting material.
For both cases we assume that there is
a good electrical contact between the F and S regions,
to assure a rather strong proximity
effect.
We summarize our results in Sec.~\ref{DiscusSection}.

\section{Model}\label{ModelSection}

The calculations of the second-order superconducting phase transition
temperature $T_c$ are based on the linearized Usadel equations
\cite{Usadel-PRL70}
for the averaged anomalous Green's functions $F_{f}$ and $F_{s}$
for the F and S regions, respectively
(see Ref.~\cite{Buzdin-RMP05} for details).
The superconducting critical temperature $T_c$ and exchange field $h$ is assumed to satisfy the dirty-limit conditions
$T_c\tau \ll 1$ and $h\tau \ll 1$, where $\tau$ is the elastic electron--scattering time.

In the F (S) region the linearized Usadel equations
take the form
\begin{eqnarray}
    & &-\frac{D_f}{2} \left(\nabla
    + \frac{2\pi i}{\Phi_0}\, \mathbf{A}\right)^2 F_f \label{eq:1} \\
    & &\qquad\qquad\qquad +\, (\,\vert\, \omega\, \vert %
               + \imath\, h\, {\rm sgn}\, \,\omega)\, F_f
    = 0\,,                                              \nonumber \\
    & &-\frac{D_s}{2} \left(\nabla
    + \frac{2\pi i}{\Phi_0}\, \mathbf{A}\right)^2 F_s
        + \vert\, \omega\, \vert F_s
    = \Delta(\mathbf{r})\,.                     \label{eq:2}
\end{eqnarray}
Here $D_{f}$  and $D_{s}$ are the diffusion constants in the ferromagnet
and superconductor, respectively,
$\omega=(2n+1)\,\pi T_c$ is a Matsubara frequency at the temperature $T_c$,
and $\Phi_0 = \pi \hbar\, c / |\,e\,|$ is the flux quantum.
Hereafter we consider only a step-like exchange field profile and, thus,
neglect both a reduction of the magnetization in the ferromagnet and
magnetization leakage into the superconductor.
Such assumption is justified above $T_c$ due to the local
nature of the exchange interaction.
Calculating the $T_c$ value itself we can assume the superconducting order parameter
 to be vanishingly small and, thus, it is natural to neglect
the spreading of the magnetic moment into the superconductor
(see, e.g.,\cite{Krivoruchko-PRB02,Bergeret-PRB04}).

The superconducting critical temperature $T_c$
is determined from the self-consistency condition for the gap function:
\begin{equation} \label{eq:3}
    \Delta(\mathbf{r})\,\ln \frac{T_c}{T_{c0}}
    + \pi T_c \sum_\omega \left(\frac{\Delta(\mathbf{r})}{\vert \omega \vert}
        - F_s(\mathbf{r},\omega)\right) = 0.
\end{equation}
Equations (\ref{eq:1}),(\ref{eq:2}) must be supplemented with
the boundary condition at the outer surfaces
\begin{equation} \label{eq:4}
     \partial_{\mathbf{n}} F_{f,s} = 0\,,
\end{equation}
and at the interface between the F and S metals:
\cite{Kuprianov-JETP88}
\begin{equation} \label{eq:5}
    \sigma_s\, \partial_{\mathbf{n}} F_s
    = \sigma_f\, \partial_{\mathbf{n}} F_f; \quad
    F_s = F_f - \gamma_b \xi_n\, \partial_{\mathbf{n}} F_f\,.
\end{equation}
Here $\xi_{s(n)}=\sqrt{D_{s(f)}/2\pi T_{c0}}$ is the superconducting
(normal-metal) coherence length,
$\sigma_{f}$ and $\sigma_{s}$ are the normal-state conductivities
of the F and S metals,
$\gamma_b$ is related to the S/F boundary resistance $R_b$ per unit area
through $\gamma_b \xi_s = R_b \sigma_f$, and $\partial_{\mathbf{n}}$
denotes a derivative taken in the direction
to the outer normal to the S surfaces, i.e., the vector
$\mathbf{n}$ is directed from the S to the F metal
at the S/F interface.
For the sake of simplicity we assume $h \gg \pi T_{c0}$.
According to the equations (\ref{eq:1}),(\ref{eq:2}),(\ref{eq:3})
there is a symmetry $F_{f,s}(\omega)=F_{f,s}^*(-\omega)$, so that
we can treat only positive $\omega$ values.

\begin{figure}
\includegraphics[width=0.45\textwidth]{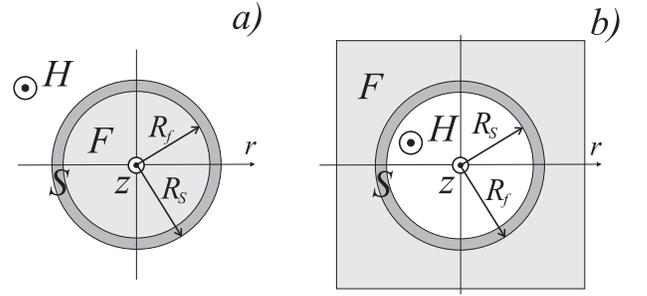}
\caption{The cross section of the hybrid S/F systems
under consideration:
a) thin-walled superconducting shell (S) around a ferromagnetic cylinder (F);
b) cylindrical cavity in a bulk ferromagnet (F) covered by a
thin layer of superconducting material (S).
Here $R_f$ is the radius of the F core (cavity), and $R_s$ is the
outer (inner) radius of the S shell; $(r,\theta,z)$ is the cylindrical
coordinate system. External magnetic field $\mathbf{H}$
applied along the $z$ axis.}
\label{Fig:1}
\end{figure}

We consider here a generic example of hybrid S/F systems
with a cylindrical symmetry:
a thin-walled superconducting hollow cylinder
placed in an electrical contact with a ferromagnet
(see Fig.~\ref{Fig:1}).
%
%
The ferromagnetic material is assumed to fill either the internal (Fig.~\ref{Fig:1}a)
or external (Fig.~\ref{Fig:1}b) region of the system.
We choose cylindrical coordinates ($r,\theta,z$) as it is
shown in Fig.~\ref{Fig:1}, take the gauge $\mathbf{A}=(0, A_\theta,0)$,
and look for a homogeneous along $z$ solution of the equations
(\ref{eq:1}),(\ref{eq:2}),(\ref{eq:3}) characterized by certain angular momentum $L$:
\begin{equation} \label{eq:6}
   \Delta(\mathbf{r})=\Delta(r)\, {\rm e}^{\imath L \theta}, \quad
   F_{f,s}(\mathbf{r})=f_{f,s}(r)\, {\rm e}^{\imath L \theta} \ .
\end{equation}
The vorticity parameter $L$ just coincides with
the angular momentum of the Cooper pair wave function.

Certainly, the magnetization $\mathbf{M}$ inside the $F$ region
makes contribution to the vector potential $\mathbf{A}$:
$$
    \mathbf{B} = \mathrm{rot}\,\mathbf{A}, \quad
    \mathbf{B} = \mathbf{H} + 4\pi \mathbf{M}\,,
$$
and can modify the conditions of superconductivity nucleation
in the S shell.
Choosing the magnetization direction along the $z-$axis
$\mathbf{M} = M \mathbf{z}_0$, we can roughly estimate
the relevant change in the total magnetic flux as follows:
 $\Phi_M \sim 4\pi^2 R_f^2 M$.
In principle, this change in the magnetic flux $\Phi_M$
modifies the period of the standard
Little-Parks oscillations and breaks the symmetry of the $T_c(H)$ dependence
with respect to the external magnetic field inversion: $T_c(H) \ne T_c(-H)$.
However, for typical parameters
$M \sim {\rm 10^2\,G}$, $T \sim {\rm 10\,K}$,
$D_s \sim {\rm 10\, cm^2/s}$ and $R_f$ of order of several
$\xi_f \sim 10\,nm$ lengths we get $\Phi_M \ll \Phi_0$,
and the asymmetry of the Little-Parks curve $T_c(H)$ may be neglected.
The above effect of the magnetization can be also
weakened provided we decrease the size of the S/F system
along the $z-$axis,
going over to the case of a thin S/F disk when the field $B$
is suppressed due to the demagnetization factor.
Note that choosing the magnetization direction in the
plane perpendicular to the  cylinder axis we can get rid
of the above magnetic flux correction completely because
of the absence of the magnetization induced field component along
the cylinder axis. These simple estimates allow us to exclude the effect of
magnetization on $T_c$ assuming that $B \simeq H$
and $A_\theta = r H / 2$.

The Usadel equations (\ref{eq:1}),(\ref{eq:2}) take the form:
\begin{eqnarray}
    & &-\frac{D_f}{2} \left[\frac{1}{r}\,\partial_r ( r\,\partial_r f_f )
        - \left(\frac{L}{r}+\frac{r}{2 a_H}\right)^2 f_f\,\right] \label{eq:7} \\
    & &\qquad\qquad\qquad +\,\imath\, h\, f_f = 0, \nonumber \\%
    & &-\frac{D_s}{2} \left[\frac{1}{r}\,\partial_r ( r\,\partial_r f_s )
        - \left(\frac{L}{r}+\frac{r}{2 a_H}\right)^2 f_s\, \right] \label{eq:8} \\
    & &\qquad\qquad\qquad +\, \omega\, f_s = \Delta\,,\nonumber %
\end{eqnarray}
where $a_H=\sqrt{\Phi_0/2\pi H}$ is the magnetic length.
An appropriate self-consistency equation (\ref{eq:3})
can be rewritten as follows:
\begin{equation} \label{eq:9}
    \Delta\,\ln\frac{T_c}{T_{c0}}
    + 2 \pi T_c \sum_{\omega>0} \left(\frac{\Delta}{\omega}
        - {\rm Re}\, f_s(\omega)\right) = 0.
\end{equation}

\section{Vortex states in thin-walled superconducting cylinder}
\label{ResultsSection}

Now we proceed with calculations of the critical temperature
dependence on the external magnetic field for two examples of multiply connected
mesoscopic hybrid S/F systems.
The first one is a ferromagnetic cylindrical filament (core) surrounded by
a thin-walled superconducting shell (see Fig.~\ref{Fig:1}a).
The second one is a cylindrical cavity in a bulk ferromagnet covered
by a thin layer of superconducting material (see Fig.~\ref{Fig:1}b).
External magnetic field $\mathbf{H}$ is assumed to be parallel
to the cylinder axis ($\mathbf{H} = H \mathbf{z}_0$)
for the both cases.

\subsection{Ferromagnetic filament covered
by a cylindrical superconducting shell}

\begin{figure}[t!]
\includegraphics[width=0.45\textwidth]{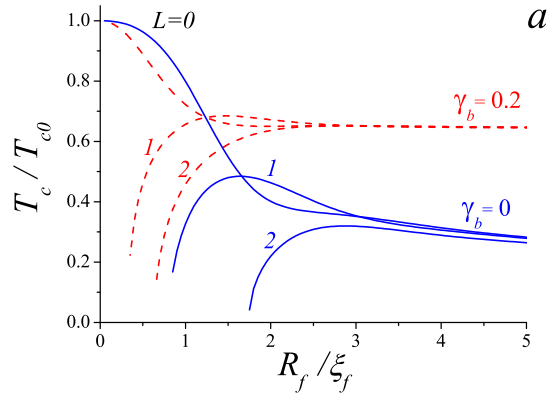}
\includegraphics[width=0.45\textwidth]{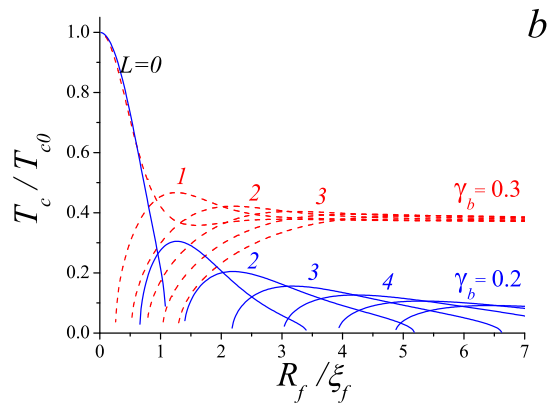}
\caption{(Color online) The typical dependences of the critical temperature $T_c$
on the F core radius $R_f$ in the absence of the external magnetic field
for different values of the interface resistance $\gamma_b$.
The numbers near the curves denote the corresponding values
of vorticity $L$. We choose the parameters
(a) $W= {\rm 0.5} \xi_s$, $\xi_s / \xi_f = {\rm 0.1}$, $\xi_n / \xi_f = {\rm 4.0}$,
$\sigma_s / \sigma_f = {\rm 1}$;
(b) $W= {\rm 0.5} \xi_s$, $\xi_s / \xi_f = {\rm 0.02}$, $\xi_n / \xi_f = {\rm 4.0}$,
$\sigma_s / \sigma_f = {\rm 0.1}$.}
\label{Fig:2}
\end{figure}

Consider a superconducting cylindrical shell of a thickness $W =
R_s - R_f\ll R_f$ surrounding a cylinder (core) of a ferromagnetic
metal.
Here $R_f$ is the radius of the F core, and $R_s$ is the outer
radius of the S shell (see Fig.~\ref{Fig:1}a). Naturally, to
observe a pronounced influence of the proximity effect on the
transition temperature, the thickness of the S shell $W$ must be
smaller than the superconducting coherence length $\xi_s$.

%
\begin{figure*}
\includegraphics[width=0.45\textwidth]{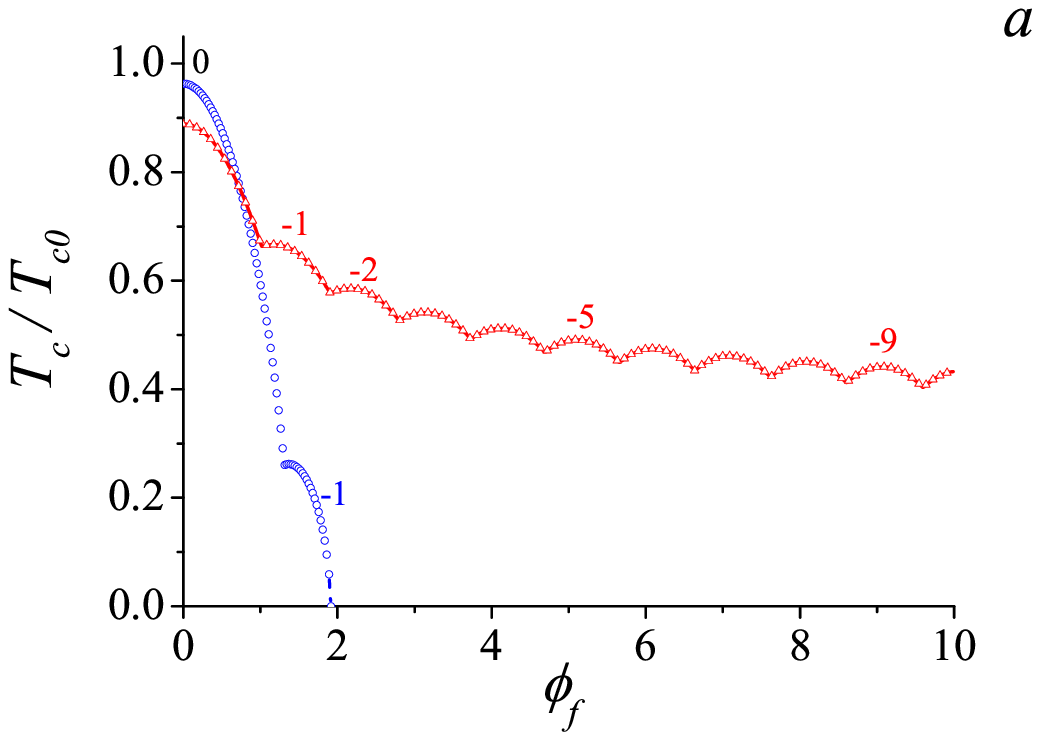}
\includegraphics[width=0.45\textwidth]{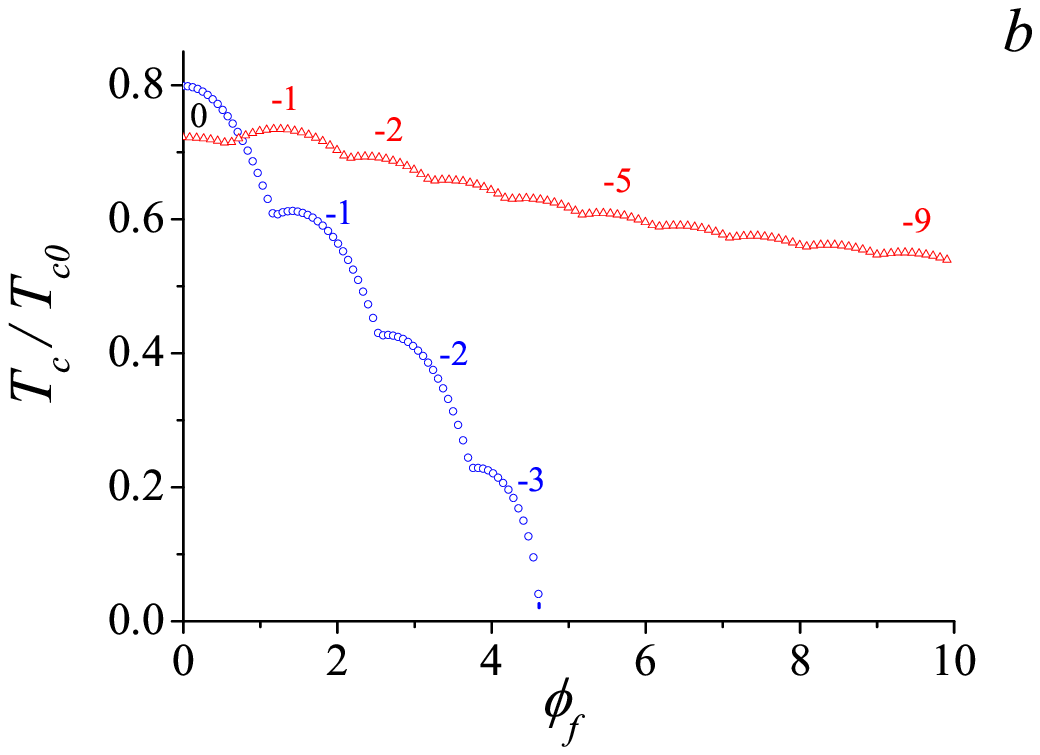}
\includegraphics[width=0.45\textwidth]{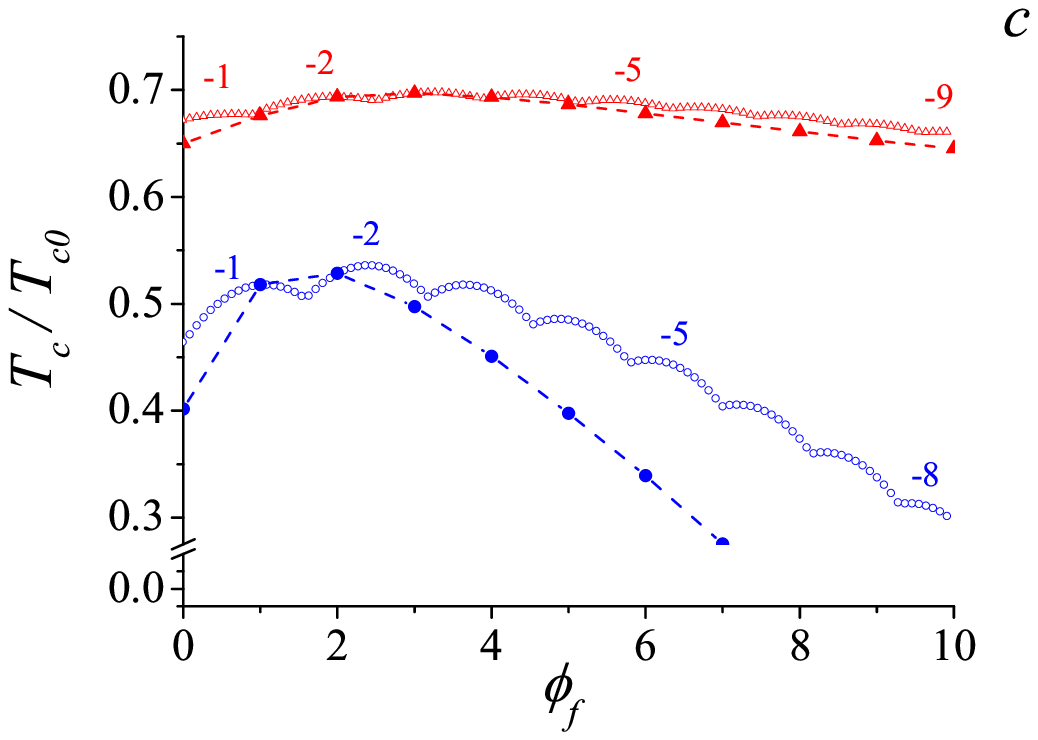}
\includegraphics[width=0.45\textwidth]{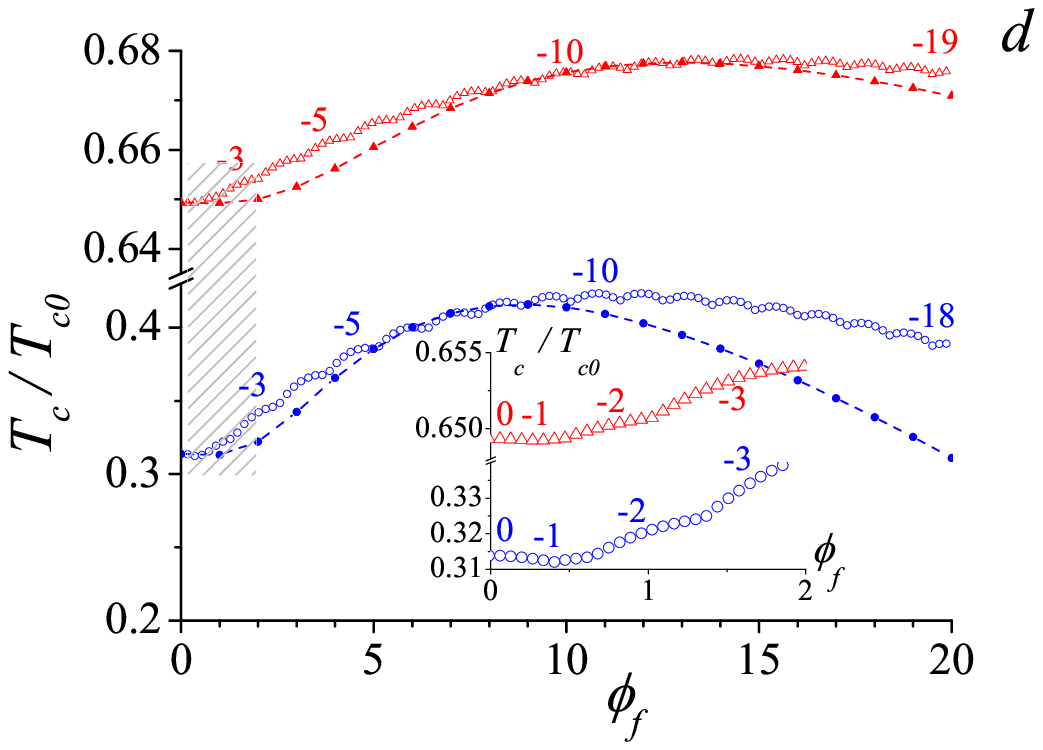}
\caption{(Color online) The typical dependences of the critical temperature $T_c$
on the external magnetic field $H$
for different values of the interface resistance
$\gamma_b$: $\gamma_b = {\rm 0}$ ($\circ$);
$\gamma_b = {\rm 0.2}$ ($\triangle$).
The magnetic field $H$ is measured in the units of
the  magnetic flux  $\phi_f$ enclosed in F cylinder.
The numbers near the curves denote the corresponding values
of vorticity $L$.
Here we choose $W= {\rm 0.5} \xi_s$;
$\xi_s / \xi_f = {\rm 0.1}$; $\xi_n / \xi_f = {\rm 4.0}$;
$\sigma_s / \sigma_f = {\rm 1}$,
and different values of the F cylinder radius
$R_f / \xi_f =$ (a) 0.5, (b) 1, (c) 2, (d) 4.
The inset in panel (d) gives the zoomed part of the $T_c(H)$ line,
marked by the shaded box.
The dashed lines in panels (c, d) are guides
for eye which connect the points corresponding to the $T_c$ values found for $\phi_f =
-L$, when the orbital effect in the depairing parameter (\ref{eq:17}) is cancelled.}
\label{Fig:3}
\end{figure*}
%

The solution of Eqn.~(\ref{eq:8}) in the F cylinder can be expressed
via the confluent hypergeometric function of the first kind
(Kummer's function) $F(a,b,z)$ \cite{Abramowitz-Handbook}
\begin{equation} \label{eq:10}
    f_f(r)= C\,\mathrm{e}^{-\phi/2} \phi^{\vert\,L\,\vert/2}
               F\left(\,a_L,\,b_L,\,\phi\,\right)\,,
\end{equation}
where $\phi$ is the flux of the external magnetic field $\mathbf{H}$
threading the circle of radius $r$ in the units
of the flux quantum $\Phi_0$
\begin{equation} \label{eq:11}
    \phi= 2 \pi r A_\theta / \Phi_0 = r^2 / 2 a_H^2\,,
\end{equation}
%
and
\begin{equation} \label{eq:12}
a_L=\frac{\vert L \vert + L +1}{2} + i{\left(\frac{a_H}{\xi_f}\right)}^2\,;
\quad
b_L=\vert L \vert +1\,.
\end{equation}
Here $\xi_f = \sqrt{D_f/h}$ is the characteristic length scale of
the order parameter variation in the F metal.
In the dirty limit, the parameter $\xi_f$ determines
both the length scale of
oscillations and the decay length for the Cooper pair wave
function in a ferromagnet.
The boundary conditions (\ref{eq:4}) and (\ref{eq:5})
for Eq.~(\ref{eq:9}) take the form
%
\begin{equation} \label{eq:13}
   \frac{d f_s}{dr}\,{\bigg\vert_{R_f}} %
      = Q_L(\phi_f)\, f_s(R_f)\,,
      \qquad \frac{d f_s}{dr}\,{\bigg\vert_{R_s}}=0\,,
\end{equation}
where $\phi_f = \pi R_f^2 H / \Phi_0$ is the flux of the external
magnetic field enclosed in the F cylinder
in the units of flux quantum $\Phi_0$, and
\begin{subequations}\label{eq:14}
\begin{eqnarray}
   & & Q_L(\phi_f) = \frac{\sigma_f / \sigma_s}
            {\gamma_b \xi_n + R_f / \kappa_L(\phi_f)}\,, \label{eq:14a} \\
   & &\kappa_L(\phi_f) = \vert L \vert - \phi_f \nonumber  \\
   & &\qquad\quad + 2\phi_f\,\frac{a_L\,F(a_L+1,\,b_L+1,\,\phi_f)}
            {b_L\,F(a_L,\,b_L,\,\phi_f)}\,. \label{eq:14b}
\end{eqnarray}
\end{subequations}
For $W \ll \xi_s$, the variations of the functions $f_s(r)$ and
$\Delta(r)$ in the superconducting shell are small: $f_s(r)\simeq
f$, $\Delta(r)\simeq \Delta$. Therefore, we can average
Eq.~(\ref{eq:8}) over the thickness of the S shell, using the
boundary conditions (\ref{eq:13}) to integrate the term
$\partial_r(r\,\partial_r f_s)$. Finally, we obtain the following
expression:
\begin{equation}\label{eq:15}
    f = \frac{\ds \Delta}{\ds \omega+\frac{D_s}{2} %
       \left[\left(\frac{L+\phi_f}{R_f}\right)^2 %
       +\frac{Q_L(\phi_f)}{W}\right]}.
\end{equation}
Substituting  the solution (\ref{eq:15}) into Eq.(\ref{eq:9})
one obtains a self-consistency equation for the critical
temperature $T_L$ of the state with a vorticity $L$:
\begin{equation}\label{eq:16}
    \ln\frac{T_L}{T_{c0}} = %
        \Psi\left(\frac{1}{2}\right)
          - Re\,\Psi\left(\frac{1}{2}+\Omega_L(\phi_f)\,\right),
\end{equation}
where $\Psi$ is the digamma function.
The depairing parameter of the mode $L$
\begin{equation}\label{eq:17}
    \Omega_L(\phi_f) = \frac{1}{2}\frac{T_{c0}}{T_L}\xi_s^2\, %
                    \left[\left(\frac{L+\phi_f}{R_f}\right)^2 %
                    +\frac{Q_L(\phi_f)}{W} \right]
\end{equation}
is responsible for the superconductivity destruction in the shell
in the applied magnetic field $\mathbf{H}$,
due to both the orbital and exchange effects.
As usual, the critical temperature $T_c$ of a superconductivity
nucleation in the shell is determined by the maximal value $T_L$:
\begin{equation} \label{eq:18}
    T_c = \underset{L}{\rm max}\{T_L\}\,.
\end{equation}
We start our numerical analysis from the case of zero external magnetic field
focusing on the effect of the S/F interface resistance $R_b$ on the
 behavior of $T_c(R_f)$.
Taking into account the asymptotic expressions for the Kummer's function
$F(a,b,z/a)$ for $\vert a \vert \to \infty$ we obtain the parameter $\kappa_L$ in a simplified form:
\begin{equation} \label{eq:19}
    \kappa_L(0)=  |\, L\, | + u_f\,\frac{I_{|L|+1}(u_f)}{I_{|L|}(u_f)}\,,
    \quad u_f = \frac{R_f}{\xi_f}(1+i)\,.
\end{equation}
The states with angular momenta $\pm L$ are degenerated for $H=0$ and have the same
critical temperature $T_c$.

In Fig.~\ref{Fig:2} we present examples of
dependencies of the critical temperature $T_c$ on the F cylinder
radius $R_f$ for  different values of the  S/F interface transparency.
We see that for a small F cylinder radius $R_f \ll \xi_f$
only the state with $L=0$ appears to be energetically favorable.
The influence of the proximity effect is weak and the critical
temperature $T_c$ is close to $T_{c0}$.
For a vortex state with $L \ge 1$ the $T_c$ value is suppressed
because of a large orbital effect.
The increase in the radius
$R_f$ results in a decrease in $T_c$ for the state with $L=0$ and
reduce the kinetic energy of supercurrents for $L \ge 1$.
At the same time, the damped-oscillatory behavior of the superconducting order
parameter in a ferromagnet becomes important, since the
diameter of the F cylinder becomes comparable with the period of the
order parameter oscillations ($\sim \xi_f$).
As a result, for $R_f > \xi_f$ the vortex state with $L=\mathrm{1}$ becomes
more energetically favorable due to the exchange interaction.
It is interesting to note that at small F cylinder radius
($R_f < \xi_f$) the critical temperature of the vortex free mode with $L=\mathrm{0}$
decreases with an increase
of the interface barrier $\gamma_b$, if the condition
$\gamma_b \xi_n R_f < \xi_f^2$ is fulfilled.
Such counterintiutive behavior is explained by the enchancement
of the pair-breaking role of the exchange field
due to the increase in the time of the Cooper pair stay
in the F metal in the presence of the interface barrier
\cite{Buzdin-RMP05}.

Figure~\ref{Fig:3} shows examples of
dependences of the critical temperature $T_c$ on the external magnetic
field $H$, obtained from Eqs.~(\ref{eq:16}), (\ref{eq:17}), (\ref{eq:18}) for
different values of F cylinder radius $R_f$.
Due to the symmetry of the phase boundary $T_c(-H)=T_c(H)$,
we present here the curves for positive values
of the external field $H$ only.
The phase boundary exhibits Little-Parks oscillations,
indicating transitions between the states with different angular momenta
$L \to L \pm 1$ of the superconducting order parameter.
For a small F cylinder radius (Fig.~\ref{Fig:3}a,b)
the influence of the exchange interaction is weak and
the $T_c(H)$ phase boundary exhibits
undamped quasiperiodic oscillations as function of magnetic field
and resembles the $T_c(H)$ curve for
a mesoscopic disk-shaped superconductor
\cite{Buisson-PL90,Jadallah-PRL99}.
It means that the S ring induces superconductivity
in the thin F filament due to proximity, and the behavior of
the S/F hybrid system under consideration
is similar to the one for a superconducting cylinder (or disk) with an inhomogeneous
order parameter.

With an increase in the F cylinder radius $R_f$ one can clearly observe
a shift of the main $T_c$ maximum towards nonzero $H$ values.
To explain this shift we note that the highest critical temperature
for a given magnetic flux $\phi_f$ corresponds to the states with the angular
momentum $L$ close to the integer part of the $-\phi_f$ value.
Exactly at the points $\phi_f=-L$ the orbital term in the depairing parameter
(\ref{eq:17}) is cancelled and the dependendence of $T_c$ vs $\phi_f$ at these points
is determined only by the exchange effect.
On the other hand, for $R_f \gtrsim \xi_f$
this exchange part of the depairing parameter is minimal and the corresponding
$T_c$ is maximal for a state with a nonzero vorticity (see Fig.~\ref{Fig:2}).
Thus, the main $T_c$ maximum shifts to a certain nonzero $\phi_f$ value.
To support this explanation we have performed simulations of the critical temperature $T_c$
for the points $\phi_f = -L$, i.e. when the orbital effect
is cancelled.
The results are shown in panels (c, d) of Fig.~\ref{Fig:3} by dashed lines.
These  curves describe a large scale (as compared to $\phi_f$)
behavior of the phase boundary $T_c(H)$
and clearly demonstrate the shift of the main $T_c$ maximum.
With the increase in the F cylinder radius $R_f$ and
the interface barrier $\gamma_b$ this main $T_c$ maximum shift
appears to increase.

Note, that a similar  shift of the main $T_c$ maximum has been observed in mesoscopic
S/F hybrid structures without the proximity effect
\cite{Lange-PRL03,Schildermans-PRB08} and in the quasi-two-dimensional organic
conductors \cite{Uji-Nature01,Balicas-PRL01}. In the latter case, an applied magnetic
field compensates the exchange fields of the paramagnetic ions and neutralizes the
destructive action of these fields \cite{Jaccarino-PRL62}.

\subsection{Cylindrical cavity in a bulk ferromagnet covered
            by a superconductor}
\begin{figure*}[t!]
\includegraphics[width=0.45\textwidth]{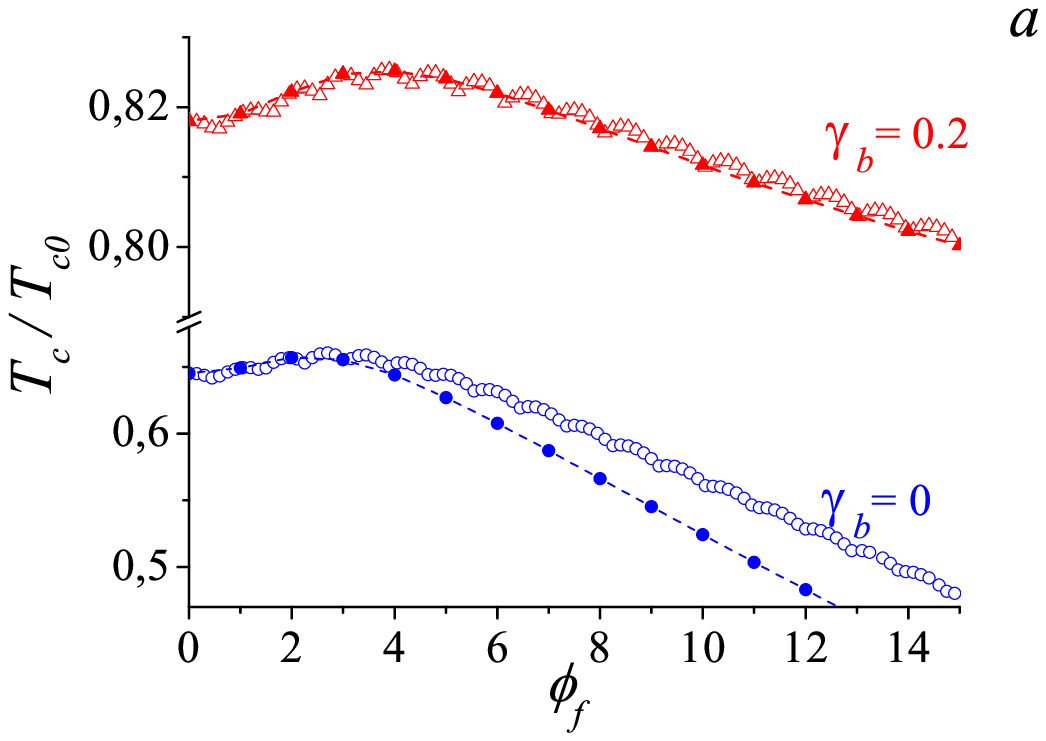}
\includegraphics[width=0.45\textwidth]{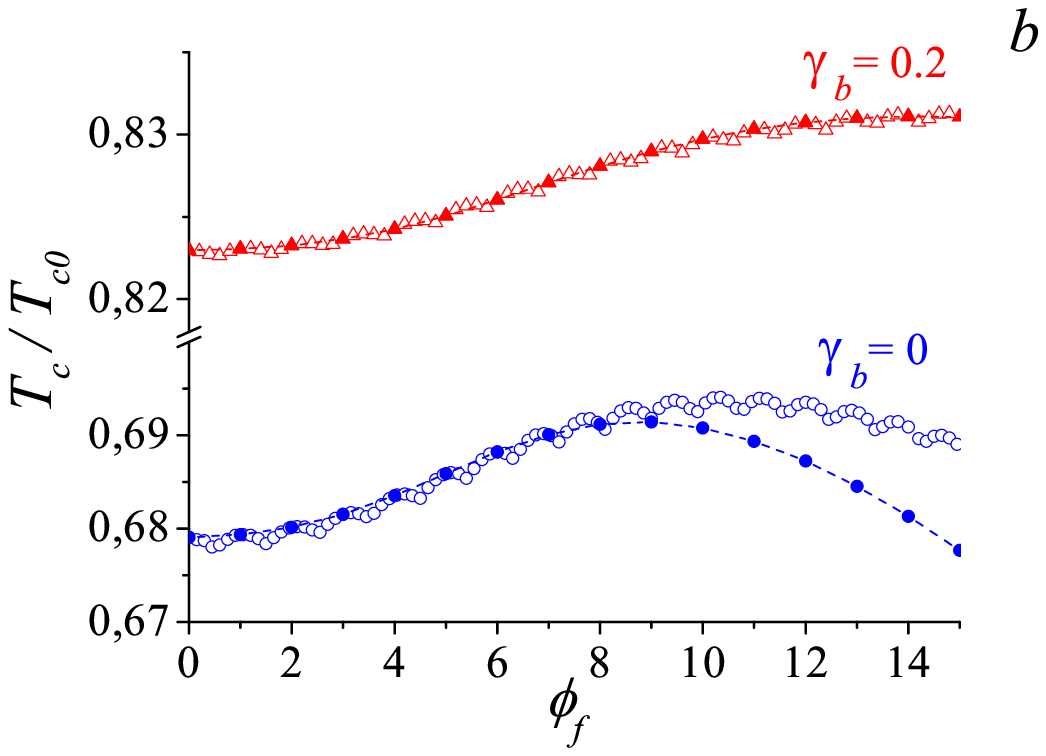}
\includegraphics[width=0.45\textwidth]{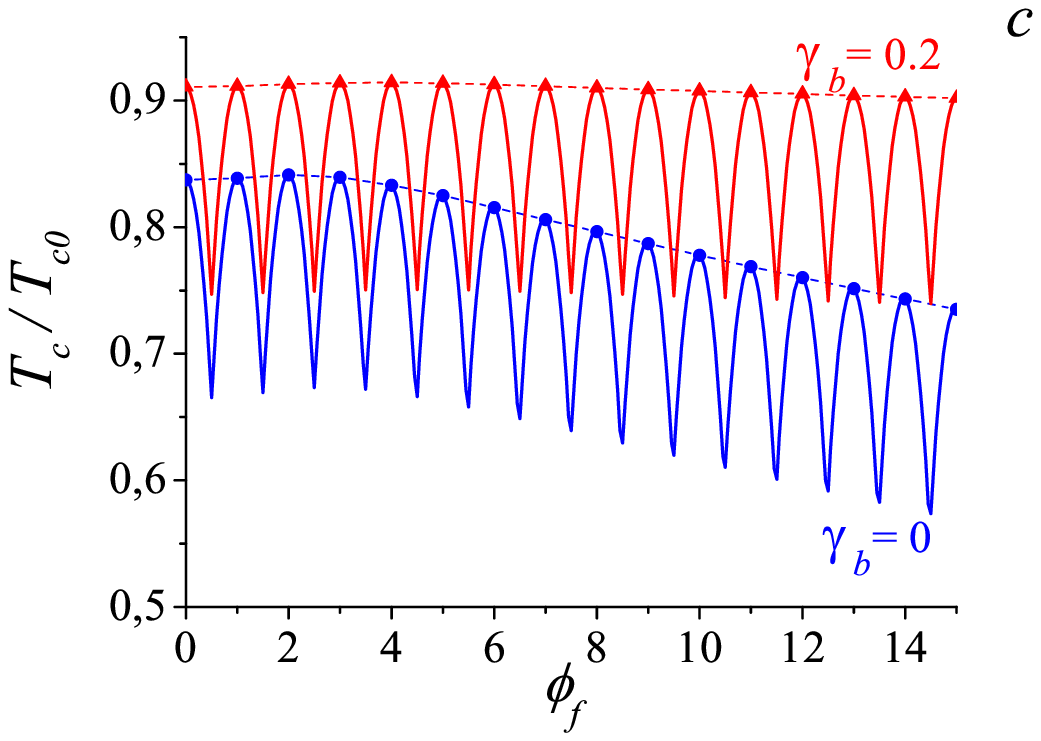}
\includegraphics[width=0.45\textwidth]{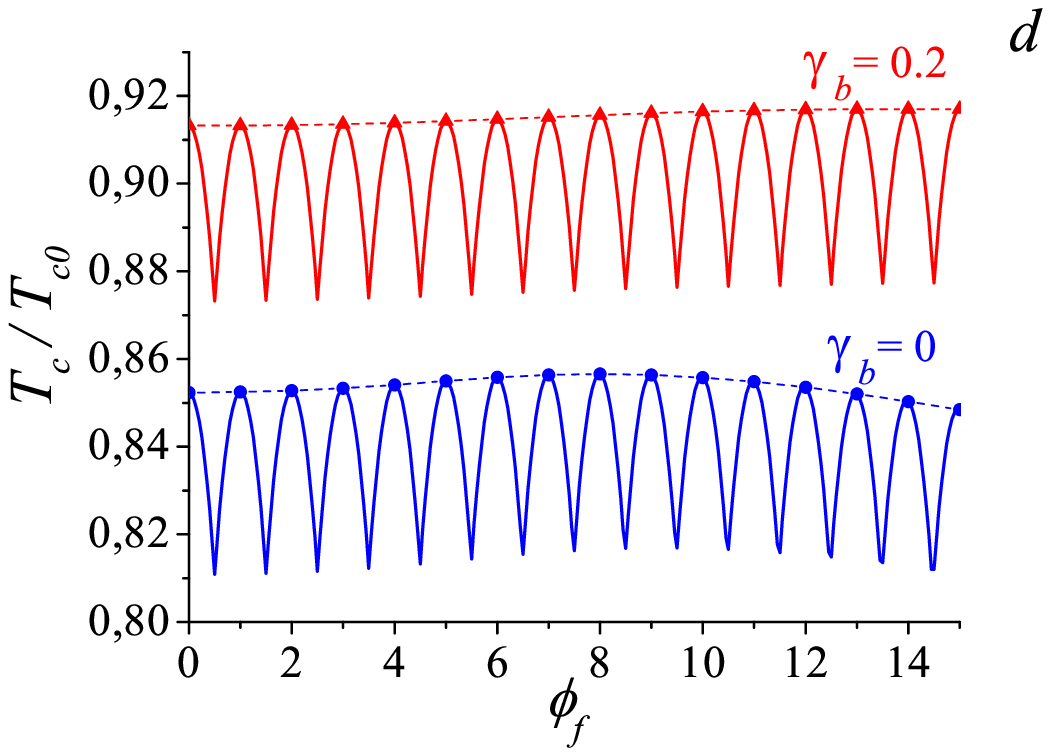}
\caption{(Color online) The typical dependencies of the critical
 temperature $T_c$ on the external magnetic field $H$
for two values of the cylindrical cavity radius $R_f$
( $R_f / \xi_f = 2$ (panels (a, c); $R_f / \xi_f = 4$ (panels b, d) )
and different values of the the interface resistance $\gamma_b$.
The magnetic field $H$ is measured in the units of
the  magnetic flux  $\phi_f$ enclosed in the cavity.
Here we choose
$W= {\rm 0.5} \xi_s$;
$\xi_n / \xi_f = {\rm 4.0}$.
The panels (a, b) show the case of the S/F system
with a rather small ratio $\xi_s / \xi_f = {\rm 0.1}$
and $\sigma_s / \sigma_f = {\rm 2}$.
The panels (c, d) show the case of the S/F system with
$\xi_s / \xi_f = {\rm 1}$ and
$\sigma_s / \sigma_f = {\rm 40}$\,.
The dashed lines are guides
for eye which connect the points corresponding to the $T_c$
values found for $\phi_f = -L$, when the orbital effect in the
depairing parameter (\ref{eq:24}) is cancelled.}
\label{Fig:4}
\end{figure*}

As a second example we consider a superconducting shell
of a thickness $W = R_f - R_s$ covering a cylindrical cavity of
the radius $R_f$
in a ferromagnetic material (see Fig.~\ref{Fig:1}b).
For simplicity, we consider only rather
thin S shells  with $W < \xi_s$ which allows us to assume
the variations of the functions $f_s(r)$ and $\Delta(r)$ in
the superconductor to be small.
The solution of Eqn.~(\ref{eq:8}) in F metal can be expressed
via the confluent hypergeometric function of the second kind
$U(a,b,z)$
\cite{Abramowitz-Handbook},
\begin{equation} \label{eq:20}
    f_f(r)= C\,\mathrm{e}^{-\phi/2} \phi^{\vert\,L\,\vert/2}
               U\left(\,a_L,\,b_L,\,\phi\,\right)\,,
\end{equation}
where the flux $\phi$, parameters $a_L$ and $b_L$ are
determined by the expressions (\ref{eq:11}) and (\ref{eq:12}).
The boundary conditions (\ref{eq:4}) and (\ref{eq:5})
for Eq.~(\ref{eq:9}) take the form
\begin{equation} \label{eq:21}
   \frac{d f_s}{dr}\,{\bigg\vert_{R_s}}=0\,,\qquad
   \frac{d f_s}{dr}\,{\bigg\vert_{R_f}} %
      = -\tilde{Q}_L(\phi_f)\, f_s(R_f)\,,
\end{equation}
where $\phi_f = \pi R_f^2 H / \Phi_0$ is the flux of the external
magnetic field enclosed in the cavity
in the units of the flux quantum $\Phi_0$,
and
\begin{subequations}\label{eq:22}
\begin{eqnarray}
   & & \tilde{Q}_L(\phi_f) = \frac{\sigma_f / \sigma_s}
            {\gamma_b \xi_n - R_f / \tilde{\kappa}_L(\phi_f)}\,, \label{eq:22a}  \\
   & &\tilde{\kappa}_L(\phi_f) = \vert L \vert - \phi_f \nonumber \\
   & &\qquad\quad - 2\phi_f\,\frac{a_L\,U(a_L+1,\,b_L+1,\,\phi_f)}
            {U(a_L,\,b_L,\,\phi_f)}\,. \label{eq:22b}
\end{eqnarray}
\end{subequations}
%
Assuming that the variations of the functions $f_s(r)$ and
$\Delta(r)$ in the superconducting shell are small ($f_s(r)\simeq
f$, $\Delta(r)\simeq \Delta$) we can average
Eq.~(\ref{eq:9}) over the thickness of the S shell, using the
boundary conditions (\ref{eq:21}) to integrate the term
$\partial_r(r\,\partial_r f_s)$.
Substitution of the solution
\begin{equation}\label{eq:23}
    f = \frac{\ds \Delta}{\ds \omega+\frac{D_s}{2} %
       \left[\left(\frac{L+\phi_f}{R_f}\right)^2 %
       +\frac{\tilde{Q}_L(\phi_f)}{W} \right]}
\end{equation}
into the self-consistency equation (\ref{eq:9})
results in the equation for the critical
temperature $T_L$ of the state with a vorticity $L$ (\ref{eq:15}),
where the depairing parameter $\Omega_L$ of the mode $L$  is
determined by the following expression:
\begin{equation}\label{eq:24}
    \Omega_L(\phi_f) = \frac{1}{2}\frac{T_{c0}}{T_L}\xi_s^2\, %
                    \left[\left(\frac{L+\phi_f}{R_f}\right)^2 %
                    +\frac{\tilde{Q}_L(\phi_f)}{W} \right]\,.
\end{equation}

Figure~\ref{Fig:4} shows examples of
dependencies of the critical temperature $T_c$ on the external magnetic field $H$,
obtained from Eqs.~(\ref{eq:16}), (\ref{eq:22}), (\ref{eq:24})
for two values of the cylindrical cavity radius $R_f$.
Similarly to the previous subsection this phase boundary resembles
 the $T_c(H)$ curve for the Little-Parks oscillations.
Slow envelopes of the oscillating curve
clearly demonstrate the shift of the main $T_c$ maximum
towards nonzero $H$ values.

\section{Summary}
\label{DiscusSection}
To sum up, we have analysed the behavior of the Little-Parks oscillations
 of the critical temperature $T_c$ on an external magnetic field $H$
 in multiply connected S/F systems affected by the exchange interaction.
As an example, we have considered mesoscopic thin-walled superconducting
cylindrical shell placed in electrical contact with a ferromagnet.
The phase-transition line $T_c(H)$ and order parameter structure have been
studied on the basis of linearized Usadel equations.
We have demonstrated that the exchange field provokes the switching between
the superconducting states with different vorticities: the interplay between
the oscillations of $T_c$ due to the Little--Parks effect and the oscillations
 due to the exchange field results in breaking of the strict periodicity of the $T_c(H)$ dependence.
We have also observed a slow modulation of the amplitude of the
quasiperiodic $T_c(H)$ oscillations.
With an increase in the superconducting shell radius the envelope of the
oscillating phase transition line $T_c(H)$ exhibits a shift of the main
 $T_c$ maximum to finite external magnetic field values.
This shift is explained by the increase in the critical temperature of the
superconducting states with nonzero vorticities due to the exchange interaction.
The above effects strongly depend on the S/F interface barrier strength.
We have shown that a decrease in the S/F transparency may stimulate the states with higher vorticities.

\section{Acknowledgments}
\label{acknowledgments}

This work was supported, in part,
by the Russian Foundation for Basic Research,
by International Exchange Program of Universite Bordeaux I,
by French ANR project "ELEC-EPR",
by the "Dynasty" Foundation,
and by the program of LEA Physique Theorique et Matiere Condensee.

\end{document}